\documentclass{article}
\usepackage{arxiv}
\usepackage{graphicx}

\usepackage{amsmath}
\usepackage{amssymb}

\usepackage[colorlinks]{hyperref} 

\usepackage[backend=bibtex,style=numeric-comp,sorting=none, doi=true, maxbibnames=9, giveninits=true,
  url=false]{biblatex}
\usepackage{booktabs}

\begin{document}




\title{Improving acoustic drone detection generalization through pretraining and data augmentation} 
\author{
Paul M. Reuter, 
Mattes Ohlenbusch, 
Christian Rollwage, 
\\
{\small Fraunhofer Institute for Digital Media Technology, Division Hearing, Speech and Audio Technology, Oldenburg, Germany}
\\
Contact: \texttt{paul.maria.reuter@idmt.fraunhofer.de}
}

\date{}
\maketitle

\begin{abstract}
Detecting unauthorized UAV flights is critical for surveillance, security, and airspace management. Acoustic drone detection, which relies on the distinctive propeller and motor sounds of UAVs, provides a low-cost, passive solution that requires no line of sight. A central challenge is generalization: reliably distinguishing drone signatures from ambient noise across unseen recording setups, environments, and UAV types (out-of-domain). Inspired by advances in large-scale audio pretraining, we develop a compact DNN-based detector and improve its generalization by (1) pretraining the model for broad sound-event classification before fine-tuning on diverse in-house and public drone recordings, and (2) applying on-the-fly augmentations (pitch shifting, noise mixing, microphone transfer function simulation, spectrogram augmentation) to expose the model to varied acoustic conditions. An ablation study quantifies the impact of each augmentation. For evaluation, we set target false-positive rates (FPR) aligned with real-world surveillance needs and report true-positive rates (TPR) on both in-domain data (public IDMT Berne 2022) and out-of-domain data (public AuDroK). Our results show that pretraining is the dominant factor for robust detection, yielding substantial TPR improvements over training from scratch on all benchmarks. The full augmentation chain provides additional gains on acoustically mismatched out-of-domain data, achieving the best mean TPR on the AuDroK subsets and the largest improvements on the most challenging scenarios. We further validate real-world applicability by measuring false positives on public non-drone corpora (IDMT-TRAFFIC and ESC-50), demonstrating equally low FPR on unfamiliar backgrounds. A distance-dependent analysis on IDMT Berne 2022 shows effective detection at distances up to 150 m.

\end{abstract}


\section{Introduction}\label{sec-intro}
In recent years, UAVs have found widespread use in cinematography, agricultural monitoring, infrastructure inspection, and military and defense applications~\cite{kaleem_amateur_2018}. 
However, easy access to UAVs presents a challenge for public security, as a single unauthorized UAV can pose a safety risk at public events or cause temporary disruption of airport traffic~\cite{sturdivant_systems_2017, al-emadi_audio_2019}.


Automatic detection of UAVs is therefore essential to enable timely countermeasures against such threats. To this end, detection systems have been proposed that utilize various sensing modalities, including radar~\cite{bertocco_malicious_2025}, video~\cite{scholes_dronesense_2022}, and audio~\cite{kang_from_2025}. In practical deployments, these modalities can be combined to improve overall performance; however, in this work we focus exclusively on acoustic drone detection.

Previous work has investigated acoustic drone detection using various features, such as short-time Fourier transform (STFT) spectrograms or Mel-frequency cepstral coefficients (MFCCs), in combination with classical machine learning methods including k-nearest neighbor classifiers~\cite{kim_real-time_2017} and support vector machines~\cite{liu_drone_2017, ohlenbusch_robust_2021, purier_uav_2024}. More recent approaches employ deep learning, training deep neural networks (DNNs) for UAV detection~\cite{kim_neural_2018, marinopoulou_two_2020} with architectures based on convolutional neural networks (CNNs) or ResNets~\cite{wang_large-scale_2022, kummritz_sound_2024}.

Although UAV detection is commonly viewed as a classification problem, UAVs are rarely encountered in practical scenarios. Consequently, acoustic drone detection systems must operate under a strict false-alarm budget: human operators can tolerate only very few false alerts over extended periods of monitoring without drone activity. At the same time, deployment conditions often differ substantially from those represented in training data, making the ability to generalize across microphones, acoustic environments, and UAV types a critical requirement.

In this work, we investigate UAV detection as a rare event classification problem, aiming at high detection rates while maintaining an acceptably low false-positive rate. To this end, we calibrate decision thresholds on background audio to achieve target false-positive rates ($\mathrm{FPR_{cal}}$) and report the resulting true-positive rate (TPR) on both in-domain and out-of-domain drone test sets. To improve generalization, we pursue two complementary strategies: large-scale AudioSet pretraining to provide transferable acoustic representations and an on-the-fly augmentation chain that exposes the model to varied acoustic conditions and reduces overfitting to dataset-specific characteristics. A systematic ablation study disentangles the contributions of each component. We further investigate the detection performance at different source–microphone distances.

\hfill
\section{Training Methodology}

In this work, we address drone detection as a binary classification problem: Given an audio segment $x_n$ with a segment index $n$, the system $D$ outputs a probability $p$ for the presence of a drone:
\begin{equation}
    p\left(\text{drone} \mid x_n\right) = D\{x_n\}.
\end{equation}
A threshold value $P_\text{drone}$ is used to decide whether a drone was detected (if $p\left(\text{drone} \mid x_n\right)\geq P_\text{drone}$). This threshold value can be calibrated after training the detection system to address trade-offs between false alarms and missed detections.

\subsection{Feature extraction}
Each input signal segment is scaled to a unit root-mean-square (RMS) value before feature extraction.
To obtain the input features for the detector, the STFT of an input signal is computed with a frame size of 25\,ms, a hop-size of 10\,ms using a Hann window. The log-Mel spectra are obtained using a Mel filterbank with 80 bands that span the frequency range [0, 8\,kHz], and then applying the base 10-logarithm to the magnitude-squared output of the filterbank. 
The resulting input features have the shape $80 \times T$ per 1\,s segment (with $T=100$ time frames).

\subsection{Detector Architecture and Pretraining}

Our detector is a compact 18-layer 2D SE-ResNet \cite{he_deep_2016} operating on log-mel spectrograms. It comprises 4 stages with filters [16, 32, 64, 128] and two Squeeze-Excitation (SE) \cite{hu_squeeze-and-excitation_2018} blocks per stage. After the final stage, we average over frequency and apply self-attentive pooling over time. A fully connected layer maps the pooled representation to a 256-dimensional vector, followed by a final linear layer that produces a single logit, which is converted to a detection probability via a sigmoid activation.

To take advantage of large-scale audio knowledge, we propose to start with a model pretrained on AudioSet~\cite{gemmeke_audioset} for general sound-event classification. The pretraining uses the same architecture and feature extraction but with 2\,s Mel-spectrogram inputs. In the fine-tuning experiments, we initialize all convolutional and dense layers of this AudioSet model and replace the classification layer with a binary output layer.

\subsection{Data Augmentation Strategies}
\label{sec:augmentations}

We use on-the-fly augmentations at both waveform and spectrogram level. At each training step, a 1\,s segment is first drawn from the combined training pool, then stochastically transformed according to configured probabilities. All augmentations preserve the 1\,s output length and are implemented in a single module that sequentially applies:

\paragraph{Drone-only pitch shifting.} For positive (drone) segments, with probability 0.25 we apply a random pitch shift of $n \in \{-5, -4, \dots, -1, 1, \dots, 5\}$ semitones using SoX's \texttt{pitch} effect. After pitch shifting, we randomly crop or pad the audio to exactly 1\,s at 16\,kHz. This exposes the model to different rotor speeds and harmonics without altering the background statistics, similarly to~\cite{kummritz_sound_2024}.

\paragraph{Background noise mixing.} With probability 0.33, we add background noise drawn from a pool of non-drone recordings (in-house array background segments, Section \ref{data_inhouse}). A random target SNR $s$  is sampled uniformly from [10, 20] dB. Denoting the clean drone signal by $x$ and the noise by $n$, with RMS amplitudes $r_x$ and $r_n$, we determine a noise gain
\[
\alpha = \frac{r_x}{r_n \cdot 10^{\,s/20}}.
\]
The scaled noise $\tilde{n} = \alpha \cdot n$ is then added to the drone signal, $x + \tilde{n}$, to obtain a mixture at the desired SNR $s$. Noise mixing was shown to improve robustness for drone detection in noisy conditions~\cite{kummritz_sound_2024}. 

\paragraph{Microphone transfer function simulation.} With probability 0.33, we apply a random infinite impulse response (IIR) filter cascade that mimics different sensor responses. The cascade consists of a first-order high-pass filter with cutoff frequency drawn uniformly from [60, 120]\,Hz, a first-order low-pass filter with cutoff in [6, 8]\,kHz, and 2–3 peaking EQ sections with random center frequency and modest gain within $\pm 4$\,dB. The cascaded filter is applied in second-order sections (SOS) form. This procedure increases variability and improves robustness to varying recording conditions and microphone transfer functions.

\paragraph{SpecAugment-style frequency masking.}
On the Mel-spectrogram level, we apply frequency masking based on SpecAugment~\cite{park2020specaugment}. With probability 0.5, we draw up to three independent frequency masks on the Mel axis. For each mask, we select a width $\Delta f$ uniformly up to 15\% of the Mel bands and a random start bin $f_0$, then set the Mel spectrogram $M[f_0 : f_0 + \Delta f, :]$ for these frequency bands to the mean of the Mel-spectrogram of the entire recording.
Time masking is not used in this work to keep the temporal envelope of drones intact. This encourages the model to rely less on narrow-band harmonic peaks and more on overall patterns, improving robustness to partial band loss (e.g., due to codecs or filtering).
\subsubsection{Ablation design}

To quantify the contribution of each component, we define a set of fine-tuning configurations:

\begin{itemize}
    \item \textbf{No-PT:} model trained from random initialization (no pretraining).
    \item \textbf{PT:} AudioSet-pretrained initialization; fine-tuned on drone data.
    \item \textbf{PT+mic:} as PT, plus microphone transfer function augmentation.
    \item \textbf{PT+mic+bg:} as PT+mic, plus background noise mixing.
    \item \textbf{PT+mic+bg+spec:} as PT+mic+bg, plus frequency masking on mel spectrograms.
    \item \textbf{PT+mic+bg+spec+pitch:} as PT+mic+bg+spec, plus drone-only pitch shifting.
\end{itemize}

Each configuration is trained and evaluated independently, enabling an ablation of pretraining and each augmentation type.

\subsection{Training Procedure}
Training is implemented in PyTorch with mixed-precision arithmetic. We use a batch size of 128 segments to compute the binary cross-entropy loss between the predicted and true class. We use the Adam optimizer with initial learning rate $10^{-3}$. Models trained from random initialization (\texttt{No-PT}) use a learning-rate decay factor of 0.9 applied every 3 epochs and are trained for 120 epochs, while fine-tuned pretrained models (\texttt{PT}) use a decay factor of 0.75 applied at every epoch and are trained for 15 epochs.
To reduce variance and enable robust comparisons, each training and augmentation configuration is repeated 10 times with different random seeds. All metrics are reported as mean $\pm$ standard deviation over these 10 runs.

\section{Datasets and Splits}
\label{sec:datasets}

We use a combination of in-house and public datasets. In the following, we distinguish (i) in-domain data recorded with our own array and drones, and (ii) out-of-domain data featuring different drones, microphones, and environments. Table~\ref{tab:datasets-overview} summarizes the role of each corpus.

\subsection{In-house array dataset (in-domain)}
\label{data_inhouse}
Our primary in-domain dataset consists of multi-channel recordings acquired with an eight-microphone array arranged in a cubic configuration with an edge length of 0.15\,m. Microphone positions are given in Table~\ref{tab:array-geometry}.

\begin{table}[t]
    \centering
    \caption{Geometry of the 8-microphone in-house array used for the in-house recordings. Positions are relative to the position of microphone 8.}
    \label{tab:array-geometry}
    \begin{tabular}{cccc}
        \toprule
        \textbf{Mic.} & \textbf{X} [m] & \textbf{Y} [m] & \textbf{Z} [m] \\
        \midrule
        1 & 0.05625 & 0.03750 & 0.15000 \\
        2 & 0.00000 & 0.09000 & 0.15000 \\
        3 & 0.01875 & 0.15000 & 0.11450 \\
        4 & 0.05625 & 0.15000 & 0.00900 \\
        5 & 0.15000 & 0.15000 & 0.07500 \\
        6 & 0.13125 & 0.07500 & 0.00900 \\
        7 & 0.15000 & 0.01875 & 0.15000 \\
        8 & 0.00000 & 0.00000 & 0.00000 \\
        \bottomrule
    \end{tabular}
\end{table}
Recordings were made in multiple sessions, where a session denotes a continuous recording campaign at a fixed site and during a specific time period, using a particular sensor deployment. In total, the dataset consists of 71\,h of drone signals and 380\,h of background signals without drones. Unless stated otherwise, single-channel signals are obtained by randomly selecting a microphone index from the array. 
To avoid overfitting to recording session-specific characteristics (e.g.\ local soundscape), we perform a session-disjoint split:
\begin{description}
    \item[IDMT-Train] Complete sessions used for training. Both drone and background segments from these sessions contribute to the training set. Background-only segments from these sessions also form part of the background pool for noise mixing (Section~\ref{sec:augmentations}).
    \item[IDMT-Test] Drone and background recordings from 8 sessions that are not used during training. This split serves as our primary in-domain test set and as the calibration source for all detection thresholds (Section~\ref{sec:evaluation-protocol}). Because both the array hardware and the drone models are shared with training, IDMT-Test evaluates in-domain detection performance.
\end{description}

\subsection{Public training data (additional drones and backgrounds)}

To increase the diversity of drones and recording conditions, we include public corpora in addition to our in-house training data :

\begin{description}
    \item[AuDroK outdoor and anechoic subsets] We use two measurement campaigns from the AuDroK collection~\cite{kummritz_comprehensive_2024}, denoted \emph{23-06-07} (outdoor) and \emph{23-02-22} (anechoic chamber). 
    Outdoor recordings were manually labeled. Anechoic chamber recordings of drone take-off and landing were excluded. 
    \item[DroneNoiseDatabase] This corpus provides drone noise recorded under field conditions~\cite{RamosRomero2023}.
    \item[SoundSnap] We draw non-drone sound effects from the commercial SoundSnap~\cite{soundsnap} library to create additional challenging background examples for the negative class. To avoid an unrealistically difficult augmentation scenario, SoundSnap files are not used as backgrounds for noise mixing with drone signals (cf.\ Section~\ref{sec:augmentations}); they appear only as stand-alone negatives.
\end{description}

\subsection{Public evaluation datasets}

To assess generalization and provide reproducible benchmarks, we employ several public datasets that are strictly excluded from training.


\subsubsection{Additional in-domain evaluation dataset (IDMT Berne~2022)}
\label{sec:distance-analysis}

For distance-dependent analysis, we also employ a public IDMT dataset~\cite{yang_neural_2025} containing drone recordings from the Berne 2022 measurement campaign. This dataset comprises real-world audio of drone flights (DJI Phantom, DJI Mavic, and Align models) recorded in August 2022 in Berne, Lower Saxony, Germany, using the \emph{same} microphone array described above. The drones perform static, linear, and random flights. We exclude the Berne2022 session entirely from training and use it solely as an additional in-domain test set, as well as to analyze detection performance as a function of source–microphone distance (Section~\ref{sec:evaluation-protocol}). 

\subsubsection{Out-of-domain evaluation datasets}

\paragraph{Out-of-domain drone datasets.} These subsets were acquired as part of the AuDroK project \cite{kummritz_comprehensive_2024} and cover different drones, microphones, and environments. In this work, we use the subsets Al-Emadi~\cite{al-emadi_audio_2019} (using only recordings with $\geq1\,$s duration), PAlloza~\cite{alloza_sound_2020}, Audacity~\cite{kummritz_comprehensive_2024}, Svanström\_Audio~\cite{svanstrom_real-time_2021}, and UBA-DLR-TUB~\cite{korper_untersuchung_2019} (referred to as AuDroK OOD).
All of these are treated as out-of-domain, because their recording sites, acquisition setups, and microphones do not overlap. While some drone models may coincide with those in the training data, any such overlap still occurs under different recording conditions.

\paragraph{Out-of-domain background datasets.} For realistic non-drone evaluation, we use:
\begin{description}
    \item[IDMT-TRAFFIC] A dataset of 17{,}506 2\,s stereo excerpts containing passing vehicles and different street-side background sounds~\cite{abeser_idmt-traffic_2021}. It covers four recording locations, four vehicle types (bus, car, motorcycle, truck), three speed-limit regimes, and dry and wet road conditions. We use all excerpts as negative (non-drone) examples. 
    \item[ESC-50] A labeled collection of 2{,}000 environmental audio recordings organized into 50 classes (e.g.\ animal sounds, human activities, natural and urban noises)~\cite{piczak_esc_2015}. We use all clips as non-drone background. 
\end{description}

\subsubsection{Segment extraction and labeling}

For all corpora, we apply the same preprocessing: First, if the corpus contains multi-channel audio, signals from a single reference channel are used. Second, audio is resampled to 16\,kHz. Third, signals are split into contiguous, non-overlapping 1\,s-segments. 
Finally, the annotation for each segment is derived from the underlying corpus annotation with either the label \emph{drone} or \emph{non-drone}, if available. For in-house recordings, labels are derived from session-level metadata.

\begin{table}[htb]
    \centering
    \caption{Overview of datasets and their roles in training and evaluation.}
    \label{tab:datasets-overview}
    \begin{tabular}{p{0.32\textwidth}ccl}
        \toprule
        \textbf{Dataset} & \textbf{Drones} & \textbf{Background} & \textbf{Role} \\
        \midrule
        IDMT-Train (in-house) & \checkmark & \checkmark & train \\
        AuDroK \textit{23-02-22} (anechoic) & \checkmark & -  & train \\
        AuDroK \textit{23-06-07} (free-field) & \checkmark & \checkmark   & train \\ 
        DroneNoiseDatabase & \checkmark & - & train \\
        SoundSnap & - & \checkmark  & train \\
        \midrule
        IDMT-Test (in-house) & \checkmark & \checkmark & in-domain test, threshold calibration \\
        IDMT Berne 2022  & \checkmark & -  & in-domain test, distance analysis \\
        \midrule
        AuDroK OOD & \checkmark & - & out-of-domain test \\
        IDMT-TRAFFIC  & - & \checkmark  & out-of-domain test \\
        ESC-50 & - & \checkmark  & out-of-domain test \\
        \bottomrule
    \end{tabular}
\end{table}

\section{Evaluation Protocol}
\label{sec:evaluation-protocol}

In our application, continuous acoustic monitoring is conducted over long periods in which drone events are relatively rare. This makes performance at low false-positive rates (FPR) more relevant than aggregate measures that average over all operating points. We therefore focus our evaluation on TPR at fixed low-FPR operating points, which directly addresses the practically relevant question: given a false-alarm budget acceptable in surveillance, how many drone events does the system detect? These operating-point–specific results are complemented by threshold-independent ROC analysis (ROC AUC) where matched positive and negative data are available.  Unless otherwise noted, we use fixed 1 s analysis windows for all experiments, reflecting the low-latency requirements of real-world drone surveillance applications. 
All metrics are computed at the segment level and reported as the mean $\pm$ standard deviation across the 10 random-seed training runs for each configuration.

\paragraph{Detection performance metrics.}

For each model, we compute scores on the IDMT-Test background segments and determine thresholds that achieve target FPRs of 1\% and 5\%. These thresholds are calibrated solely on in-domain non-drone audio and are then kept fixed for all subsequent evaluations. Using these fixed thresholds, we report the true-positive rate (denoted $\mathrm{TPR}_{\mathrm{@FPR_{cal}}=0.01}$ and $\mathrm{TPR}_{\mathrm{@FPR_{cal}}=0.05}$) on all drone test sets: IDMT-Test drones (in-domain), IDMT Berne~2022 (in-domain), and AuDroK OOD (out-of-domain). 
For AuDroK OOD, we report TPR separately for each subset as well as an aggregate TPR obtained by averaging the subset TPRs, weighting each subset equally. As a complementary, threshold-independent measure, we report ROC AUC on IDMT-Test, which is the only benchmark that contains matched drone and background recordings from the same acquisition setup.

\paragraph{Cross-domain false-positive rates.}

To verify that the calibrated operating points generalize beyond the in-domain background, we report the empirical FPR on two out-of-domain non-drone corpora (IDMT-TRAFFIC and ESC-50) at the same fixed thresholds (denoted $\mathrm{FPR}_{\mathrm{@FPR_{cal}}=0.01}$ and $\mathrm{FPR}_{\mathrm{@FPR_{cal}}=0.05}$).

\paragraph{Distance-dependent analysis.}

To study range limitations, we perform a dedicated analysis on the IDMT Berne~2022 drone recordings. The recordings are segmented into 1\,s chunks and each chunk is assigned a drone--microphone distance from the provided 3D localization metadata, binned into intervals [0--50, 50--100, 100--150, 150--200]\,m. We compute $\mathrm{TPR}_{\mathrm{@FPR_{cal}}=0.01}$ per distance bin, which yields a curve of detection probability versus range.

\section{Results and Discussion}

\subsection{Effect of pretraining and augmentation on in-domain and out-of-domain detection}

Table~\ref{tab:main-results} summarizes the main comparison across the six ablation settings. The largest effect is obtained from AudioSet pretraining. Compared to training from scratch (\texttt{No-PT}), fine-tuning a pretrained model (\texttt{PT}) improves the mean in-domain IDMT-Test $\mathrm{TPR}_{\mathrm{@FPR_{cal}}=0.01}$ from $0.860$ to $0.920$, while the corresponding ROC AUC increases from $0.972$ to $0.983$. The same trend appears on the in-domain Berne set ($\mathrm{TPR}_{\mathrm{@FPR_{cal}}=0.01}$: $0.816$ to $0.892$) and on the out-of-domain AuDroK benchmark (mean $\mathrm{TPR}_{\mathrm{@FPR_{cal}}=0.01}$: $0.701$ to $0.783$). These results show that pretraining improves both in-domain robustness and transfer to unseen drones and recording conditions.

\begin{table}[htb] 
    \centering
    \caption{Drone detection performance on in-domain and out-of-domain test sets for different pretraining and augmentation configurations, reported as ROC AUC (IDMT-Test) and $\mathrm{TPR}_{\mathrm{@FPR_{cal}}=0.01}$ (all drone test sets). All values are means $\pm$ standard deviation over 10 runs.}
    \label{tab:main-results}
    \begin{tabular}{lcccccc}
        \toprule
        & \multicolumn{2}{c}{\textbf{IDMT-Test}} & \multicolumn{1}{c}{\textbf{Berne 2022}} & \multicolumn{1}{c}{\textbf{AuDroK OOD}} \\
        \cmidrule(lr){2-3} \cmidrule(lr){4-5} \cmidrule(lr){6-7}
        \textbf{Training configuration} & AUC & $\mathrm{TPR}_{\mathrm{@FPR_{cal}}=0.01}$ & $\mathrm{TPR}_{\mathrm{@FPR_{cal}}=0.01}$ & $\mathrm{TPR}_{\mathrm{@FPR_{cal}}=0.01}$ \\
        \midrule
        No-PT & $0.972 \pm 0.002$ & $0.860 \pm 0.015$ & $0.816 \pm 0.013$ & $0.701 \pm 0.015$ \\
        PT & $0.983 \pm 0.001$ & $0.920 \pm 0.005$ & $0.892 \pm 0.007$ & $0.783 \pm 0.014$ \\
        PT+mic & $0.983 \pm 0.001$ & $0.923 \pm 0.007$ & $0.891 \pm 0.011$ & $0.816 \pm 0.012$ \\
        PT+mic+bg & $0.983 \pm 0.001$ & $0.926 \pm 0.004$ & $0.889 \pm 0.008$ & $0.818 \pm 0.010$ \\
        PT+mic+bg+spec & $0.984 \pm 0.001$ & $0.925 \pm 0.005$ & $0.890 \pm 0.007$ & $0.818 \pm 0.011$ \\
        PT+mic+bg+spec+pitch & $0.982 \pm 0.001$ & $0.921 \pm 0.004$ & $0.883 \pm 0.010$ & $0.825 \pm 0.015$ \\
        \bottomrule
    \end{tabular}
\end{table}

Augmentations provide a second, more targeted improvement. Microphone-response augmentation already yields a clear gain on AuDroK OOD, increasing the mean $\mathrm{TPR}_{\mathrm{@FPR_{cal}}=0.01}$ from $0.783$ to $0.816$. Adding background mixing further improves the in-domain operating point and yields the best IDMT-Test mean $\mathrm{TPR}_{\mathrm{@FPR_{cal}}=0.01}$ of $0.926$. Adding spectrogram masking and pitch shifting mainly benefits out-of-domain generalization: the full augmentation stack reaches the best AuDroK OOD $\mathrm{TPR}_{\mathrm{@FPR_{cal}}=0.01}$ of $0.825$. In contrast, Berne performance remains within a narrow range of $0.883$--$0.892$ for all pretrained systems, indicating that the later augmentations primarily help domain transfer rather than the in-domain test sets.

\begin{figure}[htb]
    \centering
    \includegraphics[width=1\linewidth]{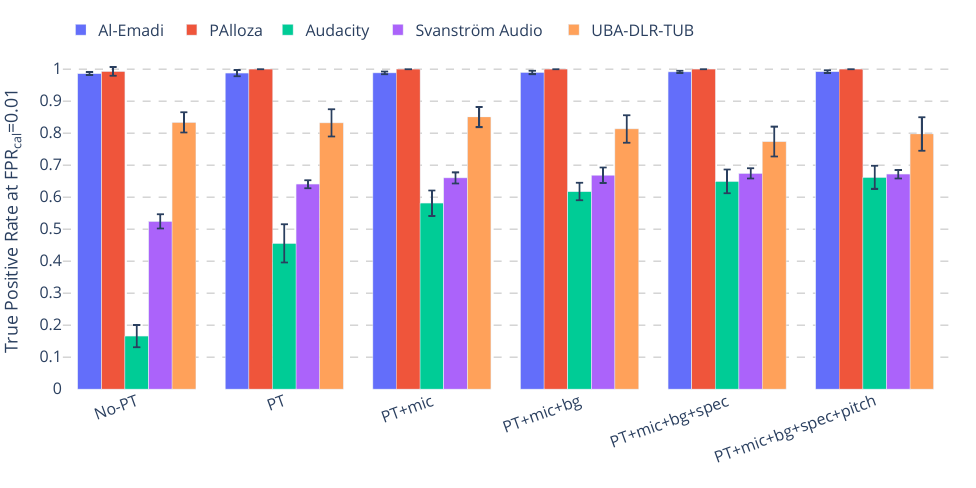}
    \caption{Subset-wise AuDroK OOD drone detection performance, reported as $\mathrm{TPR}_{\mathrm{@FPR_{cal}}=0.01}$. Bars show mean value across 10 runs and error bars indicate one standard deviation.}
    \label{fig:audrok-subsets}
\end{figure}

Because the AuDroK OOD dataset covers very different recording conditions, we report $\mathrm{TPR}_{\mathrm{@FPR_{cal}}=0.01}$ separately for each AuDroK OOD subset in Figure~\ref{fig:audrok-subsets}. 

Two subsets, Al-Emadi and PAlloza, are nearly saturated for all models ($\mathrm{TPR}_{\mathrm{@FPR_{cal}}=0.01} > 0.98$), while Audacity is clearly the hardest, with $\mathrm{TPR}_{\mathrm{@FPR_{cal}}=0.01}$ rising from $0.166$ (\texttt{No-PT}) to $0.662$ (full stack). Svanstr\"om\_Audio and UBA-DLR-TUB lie between these extremes. This confirms that the augmentation gains are not a uniform shift but concentrated on the mismatched subsets.

At the less restrictive operating point of $\mathrm{FPR_{cal}}=0.05$, the same ordering is largely preserved. 
Overall, the results support two conclusions: pretraining is the dominant contributor to the performance improvement compared to the baseline (\texttt{No-PT}), and the augmentation chain mainly improves robustness on acoustically mismatched evaluation data.

\subsection{Cross-domain false-positive rates}

Table~\ref{tab:background-transfer} summarizes the false-positive rates on IDMT-TRAFFIC and ESC-50 at the calibrated thresholds to answer the key question how well these operating points transfer to out-of-domain background corpora.

For IDMT-TRAFFIC, the transfer is very close to the calibrated operating point for all models. At the $1\%$ target, the measured FPR on IDMT-TRAFFIC remains between $0.008$ and $0.0111$; at the $5\%$ target, it stays between $0.039$ and $0.052$. ESC-50 is consistently more challenging, but the false-positive rates remain in the same low-single-digit regime: $0.019$ to $0.025$ at the $1\%$ target and $0.046$ to $0.067$ at the $5\%$ target. This shows that thresholds calibrated on the in-domain background set transfer well to another traffic-dominated background corpus and remain reasonably stable on a broader environmental sound collection.

The transfer table also indicates a mild trade-off between out-of-domain drone detection performance and background robustness. The models with the strongest AuDroK OOD performance, especially \texttt{PT+mic+bg+spec+pitch}, produce higher ESC-50 false-positive rates than models with less aggressive augmentation. However, these differences remain modest in absolute terms, and all pretrained variants operate in a practically similar low-FPR range on out-of-domain backgrounds.

\begin{table}[htb]
    \centering
    \caption{False-positive rates on out-of-domain background datasets when applying thresholds calibrated on IDMT-Test background audio to target FPRs of 0.01 and 0.05. Values are means $\pm$ standard deviation over 10 runs. 
    }
    \label{tab:background-transfer}
    \begin{tabular}{lcccc}
        \toprule
        & \multicolumn{2}{c}{\textbf{IDMT-TRAFFIC}} & \multicolumn{2}{c}{\textbf{ESC-50}} \\
        \cmidrule(lr){2-3} \cmidrule(lr){4-5} 
        \textbf{Training configuration} & $\mathrm{FPR}_{\mathrm{@FPR_{cal}}=0.01}$ & $\mathrm{FPR}_{\mathrm{@FPR_{cal}}=0.05}$ & $\mathrm{FPR}_{\mathrm{@FPR_{cal}}=0.01}$ & $\mathrm{FPR}_{\mathrm{@FPR_{cal}}=0.05}$ \\
        \midrule
        No-PT & $0.008 \pm 0.001$ & $0.043 \pm 0.006$ & $0.021 \pm 0.003$ & $0.058 \pm 0.004$ \\
        PT & $0.011 \pm 0.002$ & $0.041 \pm 0.006$ & $0.019 \pm 0.003$ & $0.046 \pm 0.006$ \\
        PT+mic & $0.010 \pm 0.002$ & $0.039 \pm 0.006$ & $0.021 \pm 0.004$ & $0.051 \pm 0.006$ \\
        PT+mic+bg & $0.010 \pm 0.002$ & $0.046 \pm 0.004$ & $0.020 \pm 0.003$ & $0.055 \pm 0.006$ \\
        PT+mic+bg+spec & $0.011 \pm 0.003$ & $0.052 \pm 0.007$ & $0.021 \pm 0.003$ & $0.058 \pm 0.004$ \\
        PT+mic+bg+spec+pitch & $0.010 \pm 0.003$ & $0.047 \pm 0.007$ & $0.025 \pm 0.002$ & $0.067 \pm 0.004$ \\
        \bottomrule
    \end{tabular}
\end{table}

\begin{figure}[htb]
    \centering
    \includegraphics[width=0.8\linewidth]{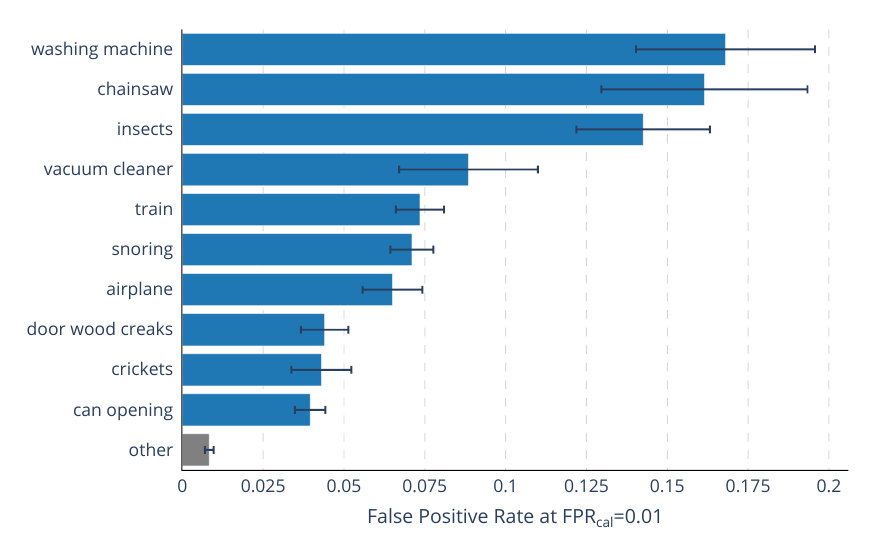}
    \caption{Class-wise false-positive rates of the full-stack model on ESC-50 at the operating point defined by $\mathrm{FPR}=0.01$ on IDMT-Test. Bars show the 10 classes with the highest mean FPR across 10 runs plus a pooled “other” category; error bars indicate one standard deviation. }
    \label{fig:esc_class_errors}
\end{figure}

A class-wise inspection of the full-stack model on ESC-50 shows that these transferred false alarms are concentrated in a limited subset of categories rather than being spread uniformly across the corpus (see Figure~\ref{fig:esc_class_errors}). The dominant confusions are \textit{washing machine}, \textit{chainsaw}, and \textit{insects}, with mean $\mathrm{FPR}_{\mathrm{@FPR_{cal}}=0.01}$ of $0.168$, $0.162$, and $0.143$, followed by \textit{vacuum cleaner} at $0.089$. This pattern is consistent with the detector reacting to sustained harmonic or quasi-stationary mechanical textures similar to those of drones.

\subsection{Distance-dependent analysis on IDMT Berne 2022 (in-domain)}

\begin{figure}[htb]
    \centering
    \includegraphics[width=0.9\linewidth]{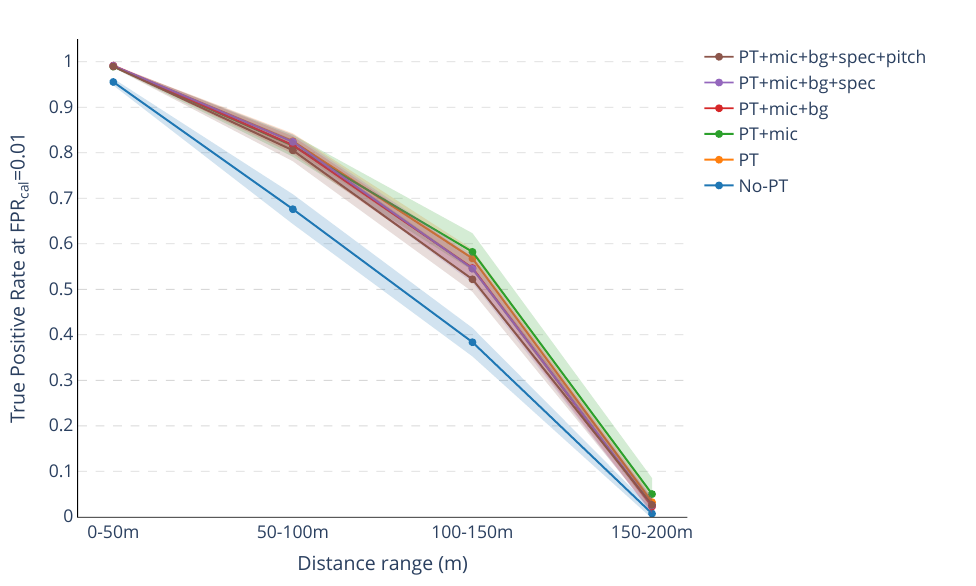}
    \caption{Detection performance on IDMT Berne 2022 as a function of drone--microphone distance, reported as $\mathrm{TPR}_{\mathrm{@FPR_{cal}}=0.01}$. Curves show mean values across 10 runs; shaded areas indicate one standard deviation.}
    \label{fig:berne-distance}
\end{figure}

Figure~\ref{fig:berne-distance} shows the Berne $\mathrm{TPR}_{\mathrm{@FPR_{cal}}=0.01}$ as a function of source distance. Pretraining substantially improves short- and medium-range performance: in the 0--50\,m range, the mean $\mathrm{TPR}_{\mathrm{@FPR_{cal}}=0.01}$ rises from $0.956$ for \texttt{No-PT} to about $0.99$ for all pretrained variants, and in the 50--100\,m range it increases from $0.676$ to approximately $0.81$--$0.83$. The most informative distance range is 100--150\,m, where the pretrained models still retain useful detection rates between $0.522$ and $0.582$, compared with only $0.384$ for \texttt{No-PT}. Beyond 150\,m, performance drops sharply for all systems, with TPR values near zero above 200\,m.

Among the pretrained models, \texttt{PT} and \texttt{PT+mic} provide the strongest medium-range performance, with \texttt{PT+mic} reaching the highest $\mathrm{TPR}_{\mathrm{@FPR_{cal}}=0.01}$ in the 100--150\,m and 150--200\,m ranges. The more aggressive augmentation stacks do not further improve long-range detection, which is consistent with the earlier observation that these augmentations mainly target cross-domain robustness rather than the already matched Berne recording setup. Overall, the distance analysis suggests that the proposed detector is highly reliable up to 100\,m, remains partially effective up to about 150\,m, and then becomes sensitivity-limited at the strict $1\%$ false-positive operating point.

\begin{figure}[htb]
    \centering
    \includegraphics[width=0.8\linewidth]{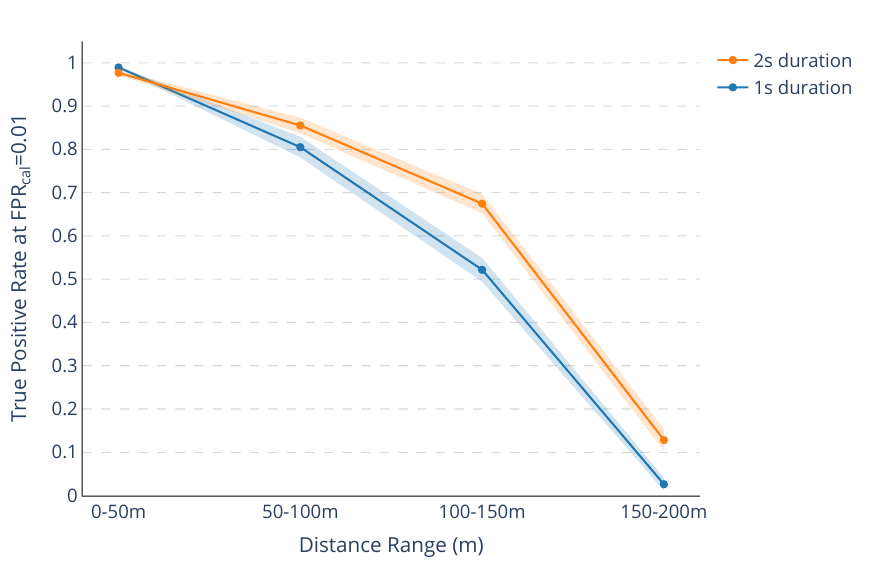}
    \caption{Detection performance on IDMT Berne 2022 as a function of drone--microphone distance for different audio durations using the full augmentation stack, reported as $\mathrm{TPR}_{\mathrm{@FPR_{cal}}=0.01}$. Curves show mean values across 10 runs; shaded areas indicate one standard deviation.}
    \label{fig:berne-distance_duration}
\end{figure}

Figure~\ref{fig:berne-distance_duration} compares the full augmentation stack for 1\,s and 2\,s test windows, where the decision threshold is calibrated independently for each duration on the corresponding IDMT-Test condition. At short range (0--50\,m), both settings are nearly saturated, with a mean $\mathrm{TPR}_{\mathrm{@FPR_{cal}}=0.01}$ of $0.990$ for 1\,s and $0.977$ for 2\,s. The benefit of the longer context becomes visible beyond 50\,m: at 50--100\,m the $\mathrm{TPR}_{\mathrm{@FPR_{cal}}=0.01}$ increases from $0.805$ to $0.855$, at 100--150\,m from $0.522$ to $0.675$, and at 150--200\,m from $0.026$ to $0.128$. This indicates that longer analysis windows are mainly beneficial once the drone becomes weaker and more temporally variable, whereas close-range detection is already saturated with 1\,s inputs.

\section{Conclusion}

In this work, we developed a compact DNN-based acoustic drone detector and systematically studied two complementary strategies to improve its generalization: large-scale AudioSet pretraining and an on-the-fly augmentation chain comprising pitch shifting, noise mixing, microphone transfer function simulation, and spectrogram masking. We evaluated the resulting models at operationally relevant false-positive rates on in-domain, out-of-domain, and distance-stratified test sets.

Our results show that AudioSet pretraining is a key factor for increasing acoustic drone detection robustness, while the proposed augmentation chain mainly improves cross-domain performance. At the primary operating point of $\mathrm{FPR_{cal}}=0.01$, the pretrained models clearly outperform training from scratch on the in-domain IDMT-Test set, on the additional Berne benchmark, and on the out-of-domain AuDroK subsets. The strongest overall out-of-domain performance is obtained with the full augmentation stack, which yields the best AuDroK mean TPR and the largest gains on the hardest subsets, especially the Audacity dataset. The large performance differences between AuDroK subsets suggest residual domain gaps that may be driven by variations in drone types, recording hardware, and signal post-processing; web-sourced material such as YouTube (e.g., the Audacity dataset) may further be affected by codec artefacts that are not yet fully captured by our current augmentations.

The distance analysis highlights an additional practical design choice: test duration matters. For the full-stack model, extending the evaluation window from 1\,s to 2\,s leaves short-range detection nearly unchanged but substantially improves medium- and long-range performance, especially beyond 100\,m. This suggests that longer temporal context can stabilize detection once the drone signal becomes weak, although it also increases latency and may reduce temporal responsiveness.

The chosen false-positive budget also has direct operational implications. 
Interpreted at the segment level, FPR$=0.01$ corresponds to roughly 36 false-positive 1\,s segments per hour, or 18 false-positive 2\,s segments per hour (if each segment is treated as an independent alarm decision). For this reason, a 1\% segment-level budget may still be too permissive for a stand-alone acoustic alarm system if every segment triggered an operator-visible alert. However, in practice, segment-level detections would typically be merged over time, filtered by additional logic, or combined with other sensing modalities. For multi-modal systems, acoustic detection can therefore operate as a sensitive early cue with a relatively relaxed threshold, whereas a stand-alone acoustic monitor would likely require stricter operating points and event-level post-processing to achieve a manageable alarm rate.

\printbibliography

@inproceedings{park2020specaugment,
  title={Specaugment on large scale datasets},
  author={Park, Daniel S and Zhang, Yu and Chiu, Chung-Cheng and Chen, Youzheng and Li, Bo and Chan, William and Le, Quoc V and Wu, Yonghui},
  booktitle={Proc. International Conference on Acoustics, Speech and Signal Processing (ICASSP)},
  pages={6879--6883},
  year={2020},
  month=may,
  doi={10.1109/ICASSP40776.2020.9053205}
}

@misc{soundsnap,
  author       = {{Soundsnap}},
  title        = {Soundsnap - Sound Effects Library},
  howpublished = {\url{soundsnap.com}},
}

@inproceedings{gemmeke_audioset,
author = {Gemmeke, Jort F. and Ellis, Daniel P. W. and Freedman, Dylan and Jansen, Aren and Lawrence, Wade and Moore, R. Channing and Plakal, Manoj and Ritter, Marvin},
title = {Audio Set: An ontology and human-labeled dataset for audio events},
year = {2017},
doi = {10.1109/ICASSP.2017.7952261},
booktitle = {Proc. International Conference on Acoustics, Speech and Signal Processing (ICASSP)},
pages = {776–780},
numpages = {5},
}

@inproceedings{hu_squeeze-and-excitation_2018,
	title = {Squeeze-and-{Excitation} {Networks}},
	doi = {10.1109/CVPR.2018.00745},
	booktitle = {Proc {IEEE} {Conference} on {Computer} {Vision} and {Pattern} {Recognition} (CVPR)},
	author = {Hu, Jie and Shen, Li and Sun, Gang},
	month = jun,
	year = {2018},
	pages = {7132--7141},
}

@inproceedings{he_deep_2016,
	title = {Deep {Residual} {Learning} for {Image} {Recognition}},
	doi = {10.1109/CVPR.2016.90},
	booktitle = {Proc. {IEEE} {Conference} on {Computer} {Vision} and {Pattern} {Recognition} ({CVPR})},
	author = {He, Kaiming and Zhang, Xiangyu and Ren, Shaoqing and Sun, Jian},
	month = jun,
	year = {2016},
	pages = {770--778},
}

@ARTICLE{kaleem_amateur_2018,
  author={Kaleem, Zeeshan and Rehmani, Mubashir Husain},
  journal={IEEE Wireless Communications}, 
  title={Amateur Drone Monitoring: State-of-the-Art Architectures, Key Enabling Technologies, and Future Research Directions}, 
  year={2018},
  volume={25},
  number={2},
  pages={150-159},
  doi={10.1109/MWC.2018.1700152}
  }

@ARTICLE{sturdivant_systems_2017,
  author={Sturdivant, Rick L. and Chong, Edwin K. P.},
  journal={IEEE Access}, 
  title={Systems Engineering Baseline Concept of a Multispectral Drone Detection Solution for Airports}, 
  year={2017},
  volume={5},
  pages={7123-7138},
  doi={10.1109/ACCESS.2017.2697979}
  }

@ARTICLE{bertocco_malicious_2025,
  author={Bertocco, Matteo and Brighente, Alessandro and Ciattaglia, Gianluca and Gambi, Ennio and Peruzzi, Giacomo and Pozzebon, Alessandro and Spinsante, Susanna},
  journal={IEEE Transactions on Instrumentation and Measurement}, 
  title={Malicious Drone Identification by Vibration Signature Measurement: A Radar-Based Approach}, 
  year={2025},
  volume={74},
  doi={10.1109/TIM.2025.3571136}
  }

@ARTICLE{scholes_dronesense_2022,
  author={Scholes, Stirling and Ruget, Alice and Mora-Martín, Germán and Zhu, Feng and Gyongy, Istvan and Leach, Jonathan},
  journal={IEEE Access}, 
  title={DroneSense: The Identification, Segmentation, and Orientation Detection of Drones via Neural Networks}, 
  year={2022},
  volume={10},
  pages={38154-38164},
  doi={10.1109/ACCESS.2022.3162866}}

@article{kang_from_2025,
    author = {Kang, Chengyang and Huang, Qinyuan and Sun, Fei and Liang, Xiuchen and Xu, Lan},
    title = {From classical approaches to recent advancements: A holistic review of acoustic detection for unmanned aerial vehicles},
    journal = {AIP Advances},
    volume = {15},
    number = {12},
    year = {2025},
    month = {12},
    doi = {10.1063/5.0304975}
}

@misc{RamosRomero2023,
  author       = {Carlos Ramos Romero and Antonio Jose Torija Martinez and Nathan Green and César Asensio},
  title        = {{DroneNoise} Database},
  month        = feb,
  year         = {2023},
  doi = {10.17866/rd.salford.22133411.v3},
  publisher    = {Zenodo},
}

@inproceedings{kummritz_comprehensive_2024,
	title = {Comprehensive {Database} of {UAV} {Sounds} for {Machine} {Learning}},
	doi = {10.61782/fa.2023.0049},
	booktitle = {Proc. {Forum} {Acusticum}},
	author = {Kümmritz, Sebastian and Paul, Lothar},
	month = jan,
	year = {2024},
	pages = {667--674},
}

@inproceedings{purier_uav_2024,
	title = {{UAV} identification from acoustic signals using statistical learning: {A} state-of-the-art},
	doi = {10.17866/rd.salford.27924897.v1},
	booktitle = {Proc. Quiet Drones},
	author = {Purier, Antoine and Bouley, Simon and Pinel-Lamotte, Lucille},
	month = sep,
	year = {2024},
}

@inproceedings{ohlenbusch_robust_2021,
	title = {Robust {Drone} {Detection} for {Acoustic} {Monitoring} {Applications}},
	doi = {10.23919/Eusipco47968.2020.9287433},
	booktitle = {Proc. {European} {Signal} {Processing} {Conference} ({EUSIPCO})},
	author = {Ohlenbusch, Mattes and Ahrens, Aike and Rollwage, Christian and Bitzer, Jörg},
	month = jan,
	year = {2021},
	pages = {6--10},
}

@inproceedings{liu_drone_2017,
	title = {Drone {Detection} {Based} on an {Audio}-{Assisted} {Camera} {Array}},
	doi = {10.1109/BigMM.2017.57},
	booktitle = {Proc. {IEEE} {International} {Conference} on {Multimedia} {Big} {Data} ({BigMM})},
	author = {Liu, Hao and Wei, Zhiqiang and Chen, Yitong and Pan, Jie and Lin, Le and Ren, Yunfang},
	month = apr,
	year = {2017},
	pages = {402--406},
}

@inproceedings{kim_real-time_2017,
	title = {Real-time {UAV} sound detection and analysis system},
	doi = {10.1109/SAS.2017.7894058},
	booktitle = {Proc. {IEEE} {Sensors} {Applications} {Symposium} ({SAS})},
	author = {Kim, Juhyun and Park, Cheonbok and Ahn, Jinwoo and Ko, Youlim and Park, Junghyun and Gallagher, John C.},
	month = mar,
	year = {2017},
}

@inproceedings{al-emadi_audio_2019,
	title = {Audio {Based} {Drone} {Detection} and {Identification} using {Deep} {Learning}},
	doi = {10.1109/IWCMC.2019.8766732},
	booktitle = {Proc. {International} {Wireless} {Communications} \& {Mobile} {Computing} {Conference} ({IWCMC})},
	author = {Al-Emadi, Sara and Al-Ali, Abdulla and Mohammad, Amr and Al-Ali, Abdulaziz},
	month = jun,
	year = {2019},
	pages = {459--464},
}

@inproceedings{abeser_idmt-traffic_2021,
	title = {{IDMT}-{Traffic}: {An} {Open} {Benchmark} {Dataset} for {Acoustic} {Traffic} {Monitoring} {Research}},
	doi = {10.23919/EUSIPCO54536.2021.9616080},
	booktitle = {Proc. {European} {Signal} {Processing} {Conference} ({EUSIPCO})},
	author = {Abeßer, Jakob and Gourishetti, Saichand and Kátai, András and Clauß, Tobias and Sharma, Prachi and Liebetrau, Judith},
	month = aug,
	year = {2021},
	pages = {551--555},
}

@inproceedings{piczak_esc_2015,
	address = {New York, NY, USA},
	title = {{ESC}: {Dataset} for {Environmental} {Sound} {Classification}},
	doi = {10.1145/2733373.2806390},
	booktitle = {Proc. {ACM} {International} {Conference} on {Multimedia}},
	author = {Piczak, Karol J.},
	month = oct,
	year = {2015},
	pages = {1015--1018},
}

@article{kim_neural_2018,
	title = {Neural {Network} based {Real}-time {UAV} {Detection} and {Analysis} by {Sound}},
	volume = {8},
	doi = {10.14801/jaitc.2018.8.1.43},
	number = {1},
	journal = {Journal of Advanced Information Technology and Convergence},
	author = {Kim, Juhyun and Kim, Dongho},
	month = jul,
	year = {2018},
	pages = {43--52},
}

@article{kummritz_sound_2024,
	title = {The {Sound} of {Surveillance}: {Enhancing} {Machine} {Learning}-{Driven} {Drone} {Detection} with {Advanced} {Acoustic} {Augmentation}},
	volume = {8},
	doi = {10.3390/drones8030105},
	number = {3},
	journal = {Drones},
	publisher = {Multidisciplinary Digital Publishing Institute},
	author = {Kümmritz, Sebastian},
	month = mar,
	year = {2024},
}

@inproceedings{svanstrom_real-time_2021,
	title = {Real-{Time} {Drone} {Detection} and {Tracking} {With} {Visible}, {Thermal} and {Acoustic} {Sensors}},
	doi = {10.1109/ICPR48806.2021.9413241},
	booktitle = {Proc {International} {Conference} on {Pattern} {Recognition} ({ICPR})},
	author = {Svanström, Fredrik and Englund, Cristofer and Alonso-Fernandez, Fernando},
	month = jan,
	year = {2021},
	pages = {7265--7272},
}

@article{korper_untersuchung_2019,
	title = {Untersuchung der {Geräuschemission} von {Drohnen} / {Investigation} of the noise emission of drones},
	volume = {14},
	doi = {10.37544/1863-4672-2019-04-10},
	number = {04},
	journal = {Lärmbekämpfung},
	author = {Körper, S. and Treichl, J.},
	year = {2019},
	pages = {108--114},
}

@inproceedings{yang_neural_2025,
	title = {Neural {Drone} {Localization} {Exploiting} {Signal} {Synthesis} of {Real}-{World} {Audio} {Data}},
	doi = {10.23919/EUSIPCO63237.2025.11226465},
	booktitle = {Proc. {European} {Signal} {Processing} {Conference} ({EUSIPCO})},
	author = {Yang, Ximei and Naylor, Patrick A. and Doclo, Simon and Bitzer, Jörg},
	month = sep,
	year = {2025},
	pages = {256--560},
}

@inproceedings{marinopoulou_two_2020,
	title = {Two {Dimensional} {Convolutional} {Neural} {Network} {Frameworks} {Using} {Acoustic} {Nodes} for {UAV} {Security} {Applications}},
	booktitle = {Proc. Quiet Drones},
	author = {Marinopoulou, Theoktisti and Vafeiadis, Anastasios and Lalas, Antonios and Rollwage, Christian and Hollosi, Danilo and Votis, Konstantinos and Tzovaras, Dimitrios},
	month = oct,
    doi = {https://doi.org/10.5281/zenodo.4543295},
	year = {2020},
}

@inproceedings{alloza_sound_2020,
	title = {Sound {Localization} of {Drones} {Using} an {Acoustic} {Camera}},
	booktitle = {Proc. Quiet Drones},
	author = {Alloza, P and Vonrhein, B and Movahed, A},
	month = oct,
	year = {2020},
}

@inproceedings{wang_large-scale_2022,
	title = {A {Large}-{Scale} {UAV} {Audio} {Dataset} and {Audio}-{Based} {UAV} {Classification} {Using} {CNN}},
	doi = {10.1109/IRC55401.2022.00039},
	booktitle = {Proc. {IEEE} {International} {Conference} on {Robotic} {Computing} ({IRC})},
	author = {Wang, Yaqin and Chu, Zhiwei and Ku, Ilmun and Smith, E. Cho and Matson, Eric T},
	month = dec,
	year = {2022},
	pages = {186--189},
}

\end{document}